**The Humanist Programming Novice as Novice**

**Ofer Elior**

*Abstract:* The primary aim of this paper is to suggest questions for future discourse and research of specialized programming courses in the Humanities. Specifically I ask whether specialized courses promote the production of fragile programming knowledge, what are the difficulties encountered by humanistic students in their learning of programming, and what may be the proper place of algorithmics in the curriculum of specialized studies.

## I. Introduction

Among scholars, educators and authors the recognition that programming skills are significant in carrying out contemporary research in the Humanities is now greater than ever. In light of this recognition, academic institutions around the world define the learning of programming as a mandatory requirement for humanistic students, sometimes as part of Digital Humanities programs, and even at the faculty level (McDaniel, 2015; Karczmarczuk, 2016; Montfort, 2016; Polefrone et al., 2016). Also, the body of scholarship about pedagogy of teaching and learning programming in the Humanities, as well as textbooks and teaching materials in this field, are constantly expanding.

An important question pertaining to these trends and one which has been a focus for educational and academic discourse, is whether an "outsourcing" of the programming education of humanistic students is a recommended practice. By this term I refer to the acquisition of fundamental programming wisdom either by enrolling to CS1 or, in some cases, by consulting programming textbooks or online courses intended to the general audience (McDaniel, 2015; Kokensparger & Peyou, 2018; Folgert et al., 2021). Particularly since the 1990s, an alternative path has been paved, as part of a broader attempt at taking control over computing education for humanistic students (Koch, 1991; Dobberstein, 1993/1994). The motivation and aim of the efforts in this direction were famously proclaimed by Koch, as follows:

> If there is to be a new, substantive area of teaching and research that combines competence in specific areas of the humanities with computer science understandings and skills, such teaching and research needs to be led by persons who themselves are competent in both the humanities and in computer science, rather than by a team of persons who represent a division of labors along the lines of 'idea' persons and 'technical' persons. (Koch, 1991, Abstract)

As far as programming studies are concerned, humanistic educators have been growingly striving to provide them to their students in their core departments, taking the wheels of shaping and molding these studies from the hands of the "technical person." To a significant extent this is due to a perception of programming studies being offered in non-humanistic frameworks, and in particular in Computer Science departments, to be unsuitable for the needs of research in the Humanities, and also as somewhat unfriendly or even boring for humanistic students (Ide, 1987; Oakman, 1987; Dobberstein, 1993/1994; Clement, 2012; Montfort, 2016; Ohman, 2019; Bleeker et al., 2022). Naturally, the move towards specialized programming studies made crucial the question of how to design these studies. Some guidelines that should be followed have been suggested and some teaching and learning practices seem to be gaining wide acceptance. However, the scope, contents and depth of the programming literacy, as well as the methodologies which should be followed in specialized studies, are still at the centre of discussions and debates (Dobberstein, 1993/1994; Clement, 2012; Bleeker et al., 2022).

The primary aim of this paper is to suggest questions for future discourse and research of specialized programming courses in the Humanities. These questions are all raised from a certain, somewhat neglected perspective, namely one which looks at the humanistic novice programmer not as humanistic but rather as novice. From this viewpoint I ask whether specialized courses promote the production of fragile programming knowledge, what are the difficulties encountered by humanistic students in their learning of programming, and what may be the proper place of algorithmics in the curriculum of specialized studies. To the best of my knowledge these questions have not hitherto been objects for empirical analysis.

## II. Literature review

Academic reports about programming studies in the Humanities, as well as programming textbooks intended for humanistic students, commonly endorse the following guideline in the design of such studies: the learning of programming should be contextualized in humanistic research. A notable implication of this guideline pertains to the choice of examples and assignments. In general programming courses students are given such exercises asking them to generate "random numbers, play blackjack with the computer, calculate cube roots, compound interest or a sphere's volume, or solve the puzzle called the Towers of Hanoi" (Sperberg-McQueen, 1987). Specialized courses do not completely overlook manipulations of numeric data (Karczmarczuk, 2016), yet they underscore the manipulation of text data and, more generally, data from humanistic disciplines (Hockey, 1986; Ide, 1987; Dobberstein, 1993/1994; Ramsay 2012; Montfort, 2016; Polefrone et al., 2016; Bleeker et al., 2022).

The use of text data is also aimed at increasing the students' motivation as much as possible. This same goal is also pursued by emphasizing the relevance and benefits of learning how to program. One pedagogic outcome of this emphasis is project-based learning (Bleeker et al., 2022). In general programming courses, it is custom to proceed, at least as far as the fundamentals of the language are concerned, by adhering to a curriculum which is, more or less, "generic", and delivered in one continuous stroke (Kokensparger, 2018). Differently in specialized studies. according to several recommendations, the learning should comprise of a set of tasks, problems or projects, a primary aim of which is to demonstrate to the students the practical uses of programming in the Humanities. The learning of the specific programming concepts required for carrying out each task is either intertwined with the work on the task or precedes it (Ide, 1987; Clement, 2012; Ramsay, 2012; McDaniel, 2015; Montfort, 2016; Birnbaum & Langmead, 2017; Kokensparger, 2018; Kokensparger & Peyou, 2018).

Another common guideline suggested for designers of specialized studies is that programming is not learned for its own sake but rather for providing humanistic students with sufficient ability and tools to carry out domain-specific tasks and to become independent learners of programming and its uses. In his *Exploratory Programming for the Arts and Humanities* Nick Montfort proclaimed this guideline as follows:

> My aim is to explain enough about programming to allow a new programmer to explore and inquire with computation, and to help such a person go on to learn more about programming while pursuing projects in the arts and humanities. (Montfort, 2016)

An important consequence of this guideline concerns the scope of discussions devoted to programming concepts. While some programming topics are deemed as significant for humanities research and thus should be included in the curriculum, others are perceived as less important and thus are either completely excluded or attended to rather briefly (McDaniel, 2015). This notion guided Montfort in his aforementioned book:

> Linked lists and binary trees are essential concepts for those learning the science of computation, but a great deal of exploration through programming can be done without understanding these concepts. Those working in artistic and humanistic areas can learn a great deal by seeing, initially, how computing allows for abstraction and generalized calculations. They can gain comfort with programming, learn to program effectively, see how to use programming as a means of inquiry – all without becoming full-blown computer scientists. For those who don't plan on getting a degree in computer science, it can sometimes be difficult to understand the bigger picture, hard to discern how to usefully compute on data and how to gain comfort with programming while also dealing with the more advanced topics that are covered in introductory programming courses. It can be hard to see the forest for the binary trees. (Montfort, 2016)

Importantly, what's left out or marginalized in specialized programming studies consists not only of advanced programming material, but also of more fundamental concepts. Thus Montfort's book, as well as Brian Kokensparger's *Guide to Programming for the Digital Humanities* (Kokensparger, 2018), both of which outline specialized studies which do not expect students to have previous programming knowledge, do not present the `while` loop, the underlying rationale probably being that the `for` loop – which they do discuss – is sufficient for most practical implementations of iterative code, and that students who wish to learn the `while` loop will be able to do that independently later on. Another example: Folgert, Kestemont and Riddel write in the Introduction to their book *Humanities Data Analysis: Case Studies with Python*, that they "do not expect the reader to have mastered the language [i.e. Python]. A relatively short introduction to programming and Python will be enough to follow along." (Folgert et al., 2021, p. 8). As an example for this kind of "short introduction" they recommend Eric Matthes' *Python Crash Course* (Matthes, 2016). The recommended book pays very little attention to certain important concepts of Python programming, e.g. sets.

Finally, the concern of providing the students an enjoying and successful learning experience has also provoked various suggestions for desirable programming languages (Koch, 1991; Ramsay, 2012; McDaniel, 2015; Karczmarczuk, 2016; Polefrone et al., 2016; Ohman, 2019). Appropriate integrated development environments, text editors, and operating systems were also discussed (Ohya, 2013; Polefrone et al., 2016).

### **III. The risk of fragile knowledge**

The first question I wish discuss is whether specialized programming courses undesirably contribute to creating fragile programming knowledge. Programming knowledge is considered to be fragile when, even if it proves to be sufficient for articulating specific notions or commands, it may be insufficient, not precise enough, and too fragmented to enable compiling a clear solution of a programming problem (Perkins, 1986). It may also make difficult the acquisition of basic skills and strategies, such as a systematic, manual tracking of a code (Gilmore, 1990; Davies, 1993; Robins, Rountree & Rountree, 2003; Lister et al., 2004). Finally, fragile programming knowledge might be inert, namely learned but remains unused (Robins, Rountree & Rountree, 2003).

Aiming at providing the know-how of humanistic programming, specialized programming courses in the Humanities have certain characteristics which might produce fragile knowledge. As said, they do not systematically teach a programming language, avoid some topics – both basic and advanced – and as far project-based studies are concerned, may involve fragmentary teaching, such that focuses only on those bits and pieces of a programming language which are considered to be sufficient for the learnt projects. In addition, teaching a programming language with the aim of only providing a limited knowledge of it, which is necessary for carrying out domain-specific tasks, may invite certain teaching practices which foster fragile knowledge, as follows.

- *Parrot-learning, or "for-doing-that-you-should-do-that" teaching.* Namely, the teacher explains how to carry out a specific procedure in a certain context, and the students are expected to replicate this operation with the same or a different data set. Following these teaching practices might place obstacles to the evolution of the students' self-identity as programmers.
- *Shallow teaching.* An inspection of recent textbooks reveals that detailed discussions of the programming concepts taught are not always, and apparently rarely are, part of specialized studies. Martin Weisser's *Python Programming for Linguistics and Digital Humanities: Applications for Text-Focused Fields* (Weisser, 2024) is a case in point. In its Introduction, Weisser presents his approach in this book as follows:

  > Other programming books may provide you with the necessary theory, walk you through code/coding examples step by step, and then give you some more advanced exercises. [...] I generally start by introducing the most essential aspects of the programming constructs covered first, [...] then ask you to apply these concepts immediately to particular questions or projects in processing textual data. (Weiser, 2024, p. 5)

  When compared to explanations in general programming textbooks, Weisser's introductions to "the most essential aspect of programming constructs" are indeed notably concise. Thus, for instance, his book devotes less than one page for discussing the `for` loop, providing no examples at all; compare e.g. Mark Lutz's bestselling *Learning Python, Fifth Edition* (Lutz, 2013), which has six pages (in a much smaller font), providing many examples.
- *Insufficient practice.* A compromise on the depth of that part of the curriculum intended for providing the fundamentals of programming might also result in not allocating enough time for practice and drill and moving too quickly to the practical uses of the acquired knowledge. This can undermine students' understanding of the learned concepts, let alone of their application, and contribute to forgetting. It should also be noted that fast pace may give rise to students' perception of the language as having a swift learning curve and, further, to anxiety or frustration (Biermann, 1998).

It may be argued that having fragile knowledge is not harmful, in the sense that it does not put obstacles to the students' programming skills and to their ability to carry out disciplinary tasks on their own after graduating from specialized courses. Following is an expression of this view made in a study about a specialized programming course of R for non-CS students, albeit not humanistic:

> [T]he parts of R programming the students will be taught should be the minimum required to handle the course. ... The important [is] that the students have heard about the concept. If they have at least some knowledge about a concept, they should be able to use the help facilities of R to look it up further." (Bååth, 1999, unpaginated)

Is this premise correct? To the best of my understanding this question was not a subject for study. Also, even if discipline-oriented teaching fosters the growth of scholars – specifically: humanistic scholars – who have the computational toolbox needed for carrying out technical research, there is the question of whether their growth is substantially encouraged in courses which attempt at reducing the risk of fragile knowledge.

## **III. General and peculiar difficulties**

General scholarship on programing education has pointed out numerous difficulties which students confront in learning various programming concepts, among them variables, conditional statements, loops, functions, recursion and so forth. Studies of these difficulties have attempted to reveal their causes and to devise learning environments, methodologies and approaches that can address them and may prove useful in minimizing their prevalence (see e.g. the surveys in: Robins, Rountree & Rountree, 2003; Medeiros, Ramalho & Falcão, 2018). Do the conclusions reached in this continuously evolving body of scholarship apply to specialized studies in the Humanities? And do students participating in these studies evince peculiar difficulties in the learning of programming concepts? Currently we can only presume the answer to the first question, given that there is no reason to assume that the answer is in the negative.

Let me illustrate a difficulty which has been examined in a study carried out in a non-Humanistic context, and may be experienced by programming students in the Humanities. My starting point is Monfort's explanation of the `for` loop. Given in the course of a discussion of a certain practical problem which necessitates iteration, his is a brief yet concise explanation, which attempts to cover all important aspects of the programming structure in question. It is formulated as follows:

> There is a kind of pattern for iteration, one that looks like this:
>
> for ____ in _____:
>
>     ____
>
> In the first blank is the variable that will be used to hold each element. It can be called almost anything—`num` is used in the example above, but it might be i or element or anything that isn't used as a Python keyword.

In the second blank is the sequence to iterate through. We might generate such a sequence with a function. We might use a sequence that is something other than a list. For now, what we have there is the list labeled `l`.

In the third blank—and possibly occupying more than one line of code, all within this same level of indentation—is whatever we want done during each iteration. This is the code that will be run "for" each element of the list. (Montfort, 2016)

Here I focus on the explication of the third blank. Montfort correctly explains that this blank represents "whatever we want done during each iteration." However the use of a *single* blank does not attend, and in fact promotes a common misunderstanding which iteration structures may present to their learners, namely the understanding that the entire sequence of actions inside the loop should be repeated. Where the loop's body includes more than one statement, some novices may follow an action-grouping misconception, thus repeating the first statement separately before repeating the subsequent action. This misconception and its reasons are discussed in a study by Grover and Basu, focusing on Computer Science students in a US middle-school. They found that a useful mean to avoid the misconception is to involve the students with activities that require them to describe what is happening in each iteration of the loop (Grover & Basu, 2017).

Another important question is to what extent difficulties with understanding programming concepts are affected by adapting explanations of such concepts in ways that presumably make them more approachable to humanistic students. As an example let us look on the code snippet in the course of its discussion Montfort explains the mechanism of the `for` loop:

```
l = [7, 4, 2, 6]
for num in l:
    print(num)
```

Designers of specialized programming studies may very well focus on adapting this code snippet to humanistic students. e.g. by scanning a list of strings instead of a list of integers. On the face of it, this adaptation does not address the said difficulty with understanding the mechanism of a loop structure.

```
names = ['Socrates', 'Plato', 'Aristotle']
for name in names:
print(name)
```

More generally, empirical studies should examine whether humanistic students who enroll to general courses encounter more difficulties than, say, Computer Science students, and also whether students in specialized programming courses in the Humanities encounter less difficulties in terms of understanding programming concepts, comparing to the difficulties they encounter in general programming courses.

### III. Promoting algorithmic thinking

The premise that programming education in the Humanities should differ from programming education in Computer Science, and the ensuing implication of narrowing the scope of these studies in terms of the taught programming concepts, seems to have led to a rather wide acceptance of the following conception: algorithmics, a meta-topic in the study of Computer Science, should not be an inherent part of specialized programming courses in the Humanities. To be sure, the advantages of teaching algorithmic thinking were not totally overlooked in the academic discourse of humanistic programming education (e.g. Polefrone et al., 2016). However currently there is at least no consensus about a need to follow practical implications of this view. Thus for example Birnbaum and Langmead proclaimed:

> Traditional computer-science concerns like data structures and algorithms and computational complexity may underlie some of what we do, but they are not typically our primary objects of study. (Birnbaum and & Langmead, 2017, p. 64)

This statement is taken as premise in some recent textbooks. For example Folgert, Kestemont and Riddel state the following in their *Humanities Data Analysis: Case Studies with Python*:

> The book is limited in that it occasionally omits detailed coverage of mathematical or algorithmic details of procedures and models, opting to focus on supporting the reader in practical work. We compensate for this shortcoming by providing references to work describing the relevant algorithms and models in "Further Reading" sections at the end of each chapter. (Folgert, Kestemont & Riddel, 2017, pp. ix-x)

In a similar vein writes Montfort in his *Exploratory Programming in Digital Humanities Pedagogy and Research*:

> In some books and courses on programming, readers learn about different sorting algorithms and about how these algorithms difer in their complexity in space and in time. These are fine topics, and necessary when building a deep foundation for those who will go on to understand the science of computation very thoroughly. If you know already that you are seeking to gain the understanding and skills equivalent to a bachelor's degree in computer science, or that you actually wish to pursue such a degree, you should probably find a more appropriate book or take a course that covers that material. (Montfort, 2016)

Kokensparger's *Guide to Programming for the Digital Humanities* is the only textbook I am aware of that shows itself to appreciate, at least in principle, the importance of developing algorithms as part of specialized programming studies. However, this conception has very few practical manifestations in the book. Additionally, similarly to other textbooks, it refers to the notion of algorithm whilst providing no explanation of what it means.

Now general scholarship on programming has time and time again emphasized that the learning of algorithmics and problem-solving is a key – and perhaps: *the* key – to good programming skills, and that ignoring these two subjects indeed harms novices' understanding and achievements (e.g. Spohrer & Soloway, 1989; Winslow 1996; McGill & Volet, 1997; Robins, Rountree & Rountree, 2003; Lister et al., 2004; Özmen & Altun, 2014). Does the fact that the audience of specialized programming courses is not Computer Science students imply that teaching algorithmics should not be given a proper place in such courses? Of course a negative answer to this question would immediately lead to another question, namely how to integrate algorithmics into the syllabuses of specialized programming courses. Currently designers of such courses are unable to easily find practical guidance. Birnbaum and Langmead, notwithstanding their general approach stated above, do provide some idea about the nature of humanistic algorithmics, yet they do so in very general terms:

> Algorithmic thinking in a humanities context means that, for example, if you want to find out which characters speak in which act of a Shakespearean play, you can ask one question in a loop over the acts instead of five separate but almost identical questions, one about each act (Birnbaum 2015). And it also means that if you want to create a word-frequency list for a text, you need to recognize that task as consisting of small subtasks, such as breaking the text into words, identifying the distinct words, counting the occurrences of each distinct word, etc. Digital humanists may someday need to know about big-O complexity and other foundations of algorithms as understood in computer science, but what humanists need to acquire immediately about algorithms is the ability to distinguish what the human does better than the computer from what the computer does better than the human, and the ability to break large, vague tasks into small, specific tasks. This requires learning to be explicit and precise in situations where humans may not otherwise have to be, but it is not computer science. (Birnbaum and & Langmead, 2017, p. 79)

To the best of my knowledge, the only substantial model currently available for embedding algorithms in teaching programming to humanistic students, can be found in one of the first textbooks, one that is a milestone in the evolution of Humanities computing education, namely Nancy M. Ide's *Pascal for the Humanities* (Ide, 1987). Its publication year and its use of mid-eighties Pascal, should not imply that this book is in any way dated with respect to the issue discussed here. Focusing on the analysis of text data, almost each of the thirteen chapters of this book deals with a certain common problem in such analysis, and at the same time continues and builds on the discussion of the problems solved in previous chapters. Perhaps as one would expect from its author – Ide is a linguistic as well as a computer scientist in her training – she puts a strong emphasis on promoting problem-solving techniques and on the development and understanding of algorithms and algorithmic notation. Thus she proclaims in the Preface to her book:

> [T]his text sets out to provide an introduction to programming that focuses on the principles of algorithmic design that underlie the programming process and to present the material in a manner suited to the humanist's habitual mode of thought. (Ide, 1987, p. x)

She then further elaborates on her method as follows:

> Following the discussion of the problem, each chapter then focuses on developing an English-language algorithm to solve it, in order to separate the process of arriving at a detailed set of steps to solve the problem from the actual implementation of these steps in a computer program. After the algorithm is fully developed, the programming language features necessary to write the program that implements the algorithm are introduced. Then the reader is taken step by step through the process of constructing the program itself, using the language features just introduced and incorporating program segments and modules from earlier chapters. This section of each chapter is typically the longest and most detailed, in order to show the reader just how algorithm is translated into program, and how individual statements and pieces of a program are fitted together to mean more than the sum of the parts. (Ide, 1987, p. xi)

Taking care to explain the meaning of "algorithm" at the beginning of the book, Ide presents algorithms with growing complexity. Thus for example this is the algorithm employed in the second chapter, "Reading and Writing Characters":

> 1. While there are more characters to be read, repeat the following:
>     a. Read a character from the input and store it in main memory.
>     b. Write a copy of the character on the display screen.

And the algorithm employed in the fourth chapter, "Counting Letters, Lines, and Sentences" (in a text file) is as follows:

> 1. Ready the file for reading.
> 2. While the character about to be read is not the end-of-file character, do the following:
>     a. While the character about to be read is not the end-of-line character, do the following:
>         (1) Read a character and store it in main memory.
>         (2) If the character stored in main memory is a letter, add 1 to the running total of the number of letters in the text.

(3) Otherwise, if the character is a period, exclamation point, or question mark, add 1 to the running total of the number of sentences in the text.
    b. Prepare to read from the beginning of the next line of input, by throwing away the carriage return that appears in the input file.
    c. Add one to the running total of the number of lines.
  3. Print the number of letters, lines, and sentences.

The algorithm to be later implemented in code is first presented by Ide in its basic form. She then gradually develop it in long and multi-stepped discussion, taking care to invite her readers to identify sub-problems and improvements needed so that the algorithm properly handle irregular scenarios. Again all this is done before coming to examine the possible code solution.

Is this intensive analysis of algorithms appropriate for the majority of humanistic beginners in programming? Sperberg-McQueen opined that that is not the case: in a review of Ide's book, she observed that it "is suited to intelligent undergraduates or more advanced students of the humanities, especially those in the text-based disciplines" (Sperberg-McQueen, 1987). Also would the students benefit from participating in a course intended mainly for algorithmic thinking prior to coming to the learning of programming? This too is a question pending future research.

## VI. Possible intermediate considerations

In the preceding I have laid out for future analysis several questions pertaining to the acquisition of programming knowledge in specialized programming courses in the Humanities. At present designers and teachers of such courses may consider some guidelines and directions which stem, I believe, from my discussion of these questions.

To begin with fragile programming knowledge, refraining from unsystematic teaching of a programming language and from fragmentary teaching seems very difficult, since these aims and practices are essentially what make specialized courses "specialized." Nonetheless, the creation of fragile knowledge may be confronted by avoiding rote learning and by encouraging the students to self-think for themselves as much as this is possible. This has already been emphasized by Kokensparger:

> [Humanistic students participating in humanistic adapted programming courses might] lose out on learning the fundamental concepts of programming. This could happen if lots of starter code and a generous number of hints are provided, to a point where the instructor is—for all intents and purposes—giving the solution code away to the students. Even a class full of DH students deserves to have a solid foundation of programming skills in an introductory

> programming course. To deny this to the class is a tragedy from both the CS and DH perspectives. (Kokensparger, 2018, p. 13)

Also in order to lessen the creation of knowledge consisting of fragile, forgotten and uncertain pieces, in-class explanations should be as deep as possible, and sufficient practice and reasonable pace must be ensured.

Consulting general scholarship on programming novices' difficulties, designers of programming courses should take action with the aim of avoiding or at least diminishing mistakes and misunderstandings which humanistic students may encounter when learning programming commands and notions. Proper attention and allocation of time must be given to the veracity and depth of such difficulties, and to ensuring adequate understanding.

Finally, designers of specialized programming courses who wish to promote their students' algorithmic thinking and problem-solving skills, may integrate into their lessons discussions of possible deconstructions of a task into sub-tasks, joint formulations of solution for each sub-task in pseudo code, and explanations of possible generalizations of the implementations of sub-tasks – all thse perhaps before writing one line of code, and contextualized in discussions of disciplinary tasks. Employing these teaching methodologies is most significant and even crucial in courses which do not suffice with parrot-teaching and require of the students to independently solve problems relying on their own thinking and not on accepted-as-given codes or code patterns. Inspecting Ide's approach and adapting it to the needs and abilities of the students is also advisable. And as mentioned, there is also the possibility of having the students come to learn programming after participating in a course intended mainly for algorithmic thinking.

## **VII. Conclusion**

The teaching of those study materials in programming which provide the basis for undertaking disciplinary tasks must not be regarded as an unavoidable preceding labor which one unfortunately must travel through in order to reach the really interesting part. Such perception invites practices which might damage the students' acquisition of programming knowledge, and by that also harm their long-run ability to independently expand their programming knowledge and apply it in their research. Also it may lead to refraining from promoting the students' algorithmic thinking, and by that avoiding an important contribution to their understanding of programming. Future research of this latter issue can be expanded to an examination of whether a place should be given in specialized courses for other skills and abilities deemed as inherent to knowledge provided in Computer Science programming courses and which I have not addressed in this paper, e.g. developing good programming habits and style, writing comments and documentation, and even emphasizing program efficiency, maintenance and style. These may not be essential for executing common tasks in humanistic computerized research. Nonetheless properly tailoring their teaching into the curriculum of specialized courses may very well contribute to achieving their primary aim. Finally, studies of difficulties which humanistic students have with understanding programming concepts will help with foreseeing and properly handling such difficulties. In a broader perspective, they will aid in listening to the voices of (humanistic) novices, voices whose consideration is part and parcel of the "technical" programming teacher's craft. And by that it will facilitate the full transformation from a traditional humanistic instructor to humanistic programming instructor.

Spohrer, J. C. & Soloway, E. (1989). "Novice Mistakes: Are the Folk Wisdoms Correct?", in E. Soloway and J. C. Spohrer (eds.), *Studying the Novice Programmer*. Hillsdale, NJ: Lawrence Erlbaum, pp. 401–416.

Weisser, M. (2024). *Python Programming for Linguistics and Digital Humanities: Applications for Text-Focused Fields*. Hoboken, NJ: John Wiley & Sons.

Winslow, L. E. (1996). "Programming Pedagogy – A Psychological Overview". *ACM SIGCSE Bulletin*, 28(3), pp. 17–22. Available at: https://doi.org/10.1145/234867.234872 .